\documentclass[epsfig,12pt]{article}
\usepackage{graphicx,epsfig}
\def\lsim{\mathrel{\raise.3ex\hbox{$<$\kern-.75em\lower1ex\hbox{$\sim$}}}}
\def\gsim{\mathrel{\raise.3ex\hbox{$>$\kern-.75em\lower1ex\hbox{$\sim$}}}}
\newcommand{ \slashchar }[1]{\setbox0=\hbox{$#1$}   
   \dimen0=\wd0                                     
   \setbox1=\hbox{/} \dimen1=\wd1                   
   \ifdim\dimen0>\dimen1                            
      \rlap{\hbox to \dimen0{\hfil/\hfil}}          
      #1                                            
   \else                                            
      \rlap{\hbox to \dimen1{\hfil$#1$\hfil}}       
      /                                             
   \fi}                                             %

\def\gev{\,{\rm GeV}}
\def\mev{\,{\rm MeV}}
\def\to{\rightarrow}
\def\re{\rm Re}

\def\r{{\cal R}}
\def\aoff{\alpha_{\rm off}}
\def\g0{{\cal G}_0}

\def\be{\begin{equation}}
\def\ee{\end{equation}}
\def\bea{\begin{eqnarray}}
\def\eea{\end{eqnarray}}
\def\bec{\begin{center}}
\def\eec{\end{center}}
\def\atversim#1#2{\lower0.7ex\vbox{\baselineskip\zatskip\lineskip\zatskip
  \lineskiplimit 0pt\ialign{$\matth#1\hfil##\hfil$\crcr#2\crcr\sim\crcr}}}

\renewcommand{\thefootnote}{\fnsymbol{footnote}}

\newcounter{appendixc}
\newcounter{subappendixc}[appendixc]
\newcounter{subsubappendixc}[subappendixc]

\renewcommand{\appendix}[1] {\vspace*{0.6cm}
        \refstepcounter{appendixc}
        \setcounter{figure}{0}
        \setcounter{table}{0}
        \setcounter{equation}{0}
        \renewcommand{\thefigure}{\Alph{appendixc}.\arabic{figure}}
        \renewcommand{\thetable}{\Alph{appendixc}.\arabic{table}}
        \renewcommand{\theappendixc}{\Alph{appendixc}}
        \renewcommand{\theequation}{\Alph{appendixc}.\arabic{equation}}
        \noindent{\bf Appendix \theappendixc #1}\par\vspace*{0.4cm}}

\begin{document}

\begin{titlepage}
\rightline{\vbox{\halign{&#\hfil\cr &KEK-TH-943 
\cr &hep-ph/0402230\cr
&2004\cr}}}
\vskip .5in
\begin{center}

{\Large\bf Scalar charmonium and glueball mixing\\  in $e^+ e^-\to J/\psi X$}

\vskip .5in \normalsize {\bf S. Dulat} \vskip .3cm Department of
Physics, Xinjiang University,
Urumqi, 830046, P.R. China \\
\vskip .3cm

\normalsize {\bf  K. Hagiwara} and {\bf  Z.-H. Lin}\\
\vskip .3cm
Theory Group, KEK, Tsukuba, Ibaraki 305-0801, Japan
\vskip 1cm

\end{center}

\begin{abstract}\normalsize
We study the possibility of the scalar charmonium and glueball
mixing in $e^+ e^-$ annihilation at $\sqrt{s}=10.6$ GeV. The
effects can be used to explain the unexpected large cross section 
($12\pm 4$ fb) and the anomalous angular distribution ($\alpha=
-1.1^{+0.8}_{-0.6}$) 
of the exclusive $e^+e^-\to J/\psi\chi_{c0}$ process observed
by Belle experiments at KEKB. We calculate the helicity amplitudes 
for the process $e^+ e^- \to J/\psi H(0^{++})$ in NRQCD, where
$H(0^{++})$ is the mixed state. We present a detailed analysis 
on the total cross section and various angular asymmetries which 
could be useful to reveal the existence of the scalar glueball state.
\end{abstract}

\vspace{1cm} PACS number(s): 12.39.Mk, 13.60.Le

\renewcommand{\thefootnote}{\arabic{footnote}}
\end{titlepage}

\section{Introduction}

The charmonium production has long been considered as a good
process for investigating both perturbative and nonperturbative
properties of quantum chromodynamics (QCD), because of the
relatively large difference between the scale at which the charm-quark
pair is produced at the parton level and the scale at which it evolves into
a quarkonium. In particular, comparing to hadron colliders, $e^+
e^-$ colliders, provide a cleaner environment to study the
charmonium productions and decays. 
However, some puzzles arise from the recent measurements on the
prompt $J/\psi$ productions at BaBar and Belle~\cite{babar, belle,
belle2}. For the inclusive $J/\psi$ productions, the cross section
is much larger than the predictions of nonrelativistic quantum
chromodynamics (NRQCD)~\cite{nrqcd}; there is also an
over-abundance of the four-charm-quark processes including
the exclusive $J/\psi$ and charmonium productions; there is no 
apparent signal in
the hard $J/\psi$ spectrum which has been predicted by the $J/\psi g g$
production mode as well as the color-octet mechanism in NRQCD. 
To provide plausible solutions and explanations for
these conflicts, theorists have studied the possibilities of the
contribution from two-virtual-photon mediate processes~\cite{bbl}, 
large higher-order QCD corrections~\cite{hagiwara,hagiwara2},
collinear suppression at the end-point region of the $J/\psi$
momentum~\cite{scet,hagiwara2}, contribution from the
$J/\psi$-glueball associated production~\cite{brodsky} and
contribution from a very light scalar boson \cite{cheung}.

Although all of the above corrections significantly enhance the
cross sections for the prompt $J/\psi$ productions, they can be
well distinguished by their momentum and angular distributions. The
two-virtual-photon processes have a very hard spectrum for the
$J/\psi$ momentum and an enhancement in the large $|\cos\theta|$
region ($\theta$ is the $J/\psi$ scattering angle in the
center-of-mass frame) due to the $t$-channel electron-exchange
contribution~\cite{bbl}. The collinear suppression softens the hard
spectrum of the color-octet production~\cite{scet} as well as that of 
the color-singlet $J/\psi gg$ process~\cite{hagiwara2}, while the
high-order QCD corrections are generally expected not to change
the angular distribution strongly and their effect could be
represented by a large renormalization $K$ factor~\cite{hagiwara2}.

$J/\psi$-glueball associated production in $e^+ e^-$ annihilation
has been shown to be of a special interest
recently~\cite{brodsky}. As a basic prediction of QCD, glueball
states have been studied in the QCD sum rules \cite{nov, nar}, bag
model \cite{jaffe, detar}, nonrelativistic potential model \cite{corn}, 
and lattice QCD \cite{bal}. In the search of the glueballs, radiative 
decays of heavy quarkonium, such as $J/\psi \to \gamma X$ and $\Upsilon 
\to \gamma X$, are regarded as favorable processes for glueball 
production and high statistics data have been studied by BES and CLEO
Collaborations~\cite{bes,cleo}. The quarkonium-glueball
associated productions at B factories should be another good place to
discover glueballs because their amplitudes are essentially the same
as those of the quarkonium decays. If glueball masses lie in the
charmonium mass region, the glueball may be hidden in the charmonium
resonance peak in the exclusive production, and this mechanism may contribute 
to the excess of exclusive charmonium-pair production events. In
Ref.~\cite{brodsky}, $J/\psi$-glueball productions 
$e^+ e^- \to J/\psi gg \to J/\psi \cal{G}$ have been studied, and at the
leading-twist level and the leading-order of the NRQCD
velocity-power-counting rule, the production of the scalar
glueball state $\g0(0^{++})$ is found to be dominant. As we shall see, the
angular distribution of $e^+ e^- \to J/\psi \g0$ is dramatically
different from those of the other exclusive $J/\psi$-charmonium productions, 
such as $e^+ e^- \to J/\psi \eta_c$ and $e^+ e^- \to J/\psi \chi_{c0}$. The
angular distributions of all those processes with single virtual photon
exchange can be parametrized as
\bea\label{angular}
\frac{d\sigma} {d \cos\theta} \sim 1+ \alpha \cos^2\theta, 
\eea
by using an asymmetry parameter $\alpha$ ($-1\leq\alpha\leq 1$).
For $e^+ e^- \to J/\psi \eta_c$ and $e^+ e^- \to J/\psi
\chi_{c0}$, $\alpha=1$ and 0.25 respectively \cite{braaten}, 
while for the $J/\psi$-$\g0$ production mode, $\alpha$ is estimated 
to be $-$0.85 in NRQCD \cite{brodsky}. We note here
that preliminary result on the $J/\psi$ angular distribution in the 
$J/\psi$-$\chi_{c0}$ region is $\alpha=-1.1^{+0.8}_{-0.6}$, which is 
more consistent with $\alpha\sim -1$ rather than $\alpha\sim 0.25$
~\cite{belle2}. Conservatively, the present Belle data~\cite{belle2} 
suggest that $\alpha$ should be negative. This unexpected angular 
distribution leads us to consider further the possibility of the scalar
charmonium $\chi_{c0}$ and glueball $\g0$ mixing. In this Letter,
we will focus on the mixing effects on various angular distributions which
could be useful to reveal the existence of the scalar glueball state.

\section{Formulae}
Let us start with the angular momentum and parity considerations. The
$J/\psi$-scalar($H$) associated production via a virtual photon in
the $J^{PC}$ notation is $1^{--} \to 1^{--}+ 0^{++}$. If we denote by $L$
the orbital angular momentum between $J/\psi$ and $H$, the angular momentum
conservation tells $L=0,1,2$. Conservation of parity gives 
$-1=1\times(-1)\times(-1)^L$, and hence $L$ should be even. We are left with 
$L=0$ or 2, that is, only $S$-wave and $D$-wave $J/\psi$-$H$ are allowed.

Now we present the helicity amplitudes for the $J/\psi$-$\chi_{c0}$ and 
$J/\psi$-$\g0$ productions separately in NRQCD. One of the crossed Feynman 
diagrams for $e^+ e^- \to J/\psi \chi_{c0}$ is shown in Fig. 1(a), and that
for $e^+ e^- \to J/\psi \g0$ is shown in Fig. 1(b). We adopt the standard 
covariant projection formalism for the color singlet productions
\cite{kuhn}. If a charmonium carries the momentum $P$, 
then the charm quark and anti-quark within the
charmonium are on-shell and carry momenta $P/2-p$ and $P/2+p$,
where $p$ is the relative momentum of the quark pair and satisfies
$p\cdot P=0$. The amplitudes for the charm-pair production are
furthermore projected into a certain spin component by the
projection operator \bea {\cal P}_{S,S_z}=\sum_{s1,s2}
v(P/2-p;s_2){\bar u}(P/2+p;s_1) \langle 1/2,s_1; 1/2,s_2|S,S_z
\rangle. \eea After expanding the amplitudes in $p$ and convoluting
them with the wave functions at the origin, one obtains the
helicity amplitudes for $S$-wave and $P$-wave charmonia, such as
$J/\psi$ and $\chi_{c0}$. For the glueball production as shown in
Fig. 1(b), we follow the method of Ref.~\cite{brodsky} and 
extract the leading-twist contribution by expanding
the gluon momenta in the light-cone direction and requiring the
gluon pair to be collinear. We define the light-cone direction as
$n^\mu=(1,0,0,1)$, while the $J/\psi$ momentum direction is along
$\bar{n}=(1,0,0,-1)$. The formation of the scalar glueball $\g0$
from the gluon pair is described by a collinear wave function
$I_0=\int^1_0 dx\phi_0(x)$, where $\phi_0$
is the distribution wave function for $\g0$ and $x$ is the
light-cone momentum fraction. We denote the $\g0$ momentum by
$k=(k^+=k\cdot {\bar n}, k^-=k\cdot n,{\bf 0_\perp})$, 
and further represent the transverse 
momentum of a gluon inside $\g0$ as ${\bf k_\perp}$.
By integrating over $\bf{k_\perp}$ and transforming from the momentum 
to the coordinate space, the distribution wave function $\phi_0(x)$ 
is obtained as
\bea
\phi_0(x)=\frac{F^0_{\alpha\beta}}{\sqrt{2(N_c^2-1)}}
\int\frac{d^2{\bf k_\perp} dz^+ d^2{\bf z_\perp}}{(2\pi)^3 k^-
x(1-x)} e^{-i(x k^- z^+ -{\bf k_\perp}\cdot {\bf z_\perp})}
\langle \g0 | T G^{- \alpha}_a(0^-,z^+,{\bf z_\perp}) G^{-
\beta}_a (0)|0 \rangle. 
\eea 
Here $F^0_{\alpha\beta}= [-g_{\alpha\beta}+(n_\alpha
{\bar n}_\beta+{\bar n}_\alpha n_\beta)/2]/\sqrt{2}$ 
is the scalar projector. $G^-_a$ is the gluon field
along the light-cone direction and $a$ is the color index.

\begin{figure}
\centerline{\epsfysize 1.5 truein \epsfbox{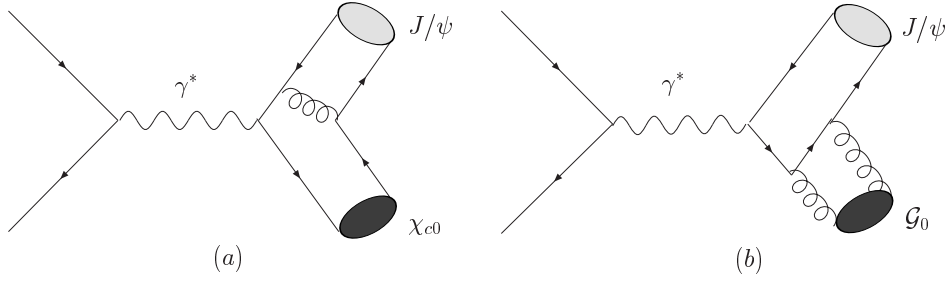}}
\caption{
(a) One of the four Feynman diagrams for $e^+ e^- \to J/\psi \chi_{c0}$;
(b) One of the six Feynman diagrams for $e^+ e^- \to J/\psi \g0$.
}
\end{figure}

Under the above approximations, the helicity amplitudes
for $J/\psi$-$\chi_{c0}$ production are given by,
\begin{eqnarray}
M_\chi(\sigma, \lambda) = g^2_s e^2 Q_c
\frac{16}{\sqrt{3} s^4 m_c}
           \Big[ (9 s^2 - 56 s m^2_c)  j_\sigma \cdotp\epsilon^*_\lambda
              + (-14s + 24 m^2_c)    j_\sigma\cdotp P_1
               \epsilon^*_\lambda \cdotp P_2
           \Big] \Phi(0) \Phi'(0),
\end{eqnarray}
and those for $J/\psi$-$\g0$ production are
\begin{eqnarray}
M_g(\sigma, \lambda) = g^2_s e^2 Q_c \frac{16 \sqrt{m_c}}{s}
                   \Big[\frac{ j_\sigma\cdotp \epsilon^*_\lambda}{ s - 4 m^2_c}
                        - \frac{ j_\sigma \cdotp n
                \epsilon^*_\lambda\cdotp n}{2 s}\Big ]
                  \Phi(0) I_0,
\end{eqnarray}
where $j$ is the electron-positron current and $\epsilon$ is the 
polarization vector of $J/\psi$. $\sigma$ and $\lambda$ denote the 
helicities of the electron-positron current (along the electron beam 
direction) and $J/\psi$, respectively.
$\Phi(0)$ and $\Phi'(0)$ are the wave functions at the origin
for $J/\psi$ and $\chi_{c0}$.  
In the center-of-mass frame, the amplitudes are reduced to
\begin{eqnarray}\label{hel1}
M_{\chi,g}(\sigma = \pm, \lambda = 0) &=& g^2_s e^2 Q_c
a_{\chi,g}
 \sin\theta, \nonumber \\
M_{\chi,g}(\sigma = \pm, \lambda = \pm)&=& g^2_s    e^2    Q_c
  b_{\chi,g}   \frac{\sigma + \lambda    \cos\theta}{\sqrt{2}},
\end{eqnarray}
where the coefficients $a_{\chi,g}$ and $b_{\chi,g}$ are for the
longitudinally and transversely polarized $J/\psi$ amplitudes respectively.
We find
\begin{eqnarray}\label{abchi}
a_\chi &=& C_F \frac{32}{\sqrt{3}} \frac{1}{s^2 r^2}
   ( 4+ 10 r^2 - 3r^4 ) \Phi(0)
\Phi'(0),
\nonumber\\
b_\chi &=& C_F \frac{32}{\sqrt{3}}   \frac{1}{s^2 r}
   (-18 + 7 r^2   ) \Phi(0) \Phi'(0),
\end{eqnarray}
for $J/\psi$-$\chi_{c0}$ production, and  
\begin{eqnarray}\label{abg}
a_g &=& \sqrt{C_F} \frac{32}{s \sqrt{m_c}}\frac{1}{4-r^2}
                \Phi(0)   I_0, \nonumber\\
b_g &=& -\sqrt{C_F}   \frac{16}{s\sqrt{m_c}} \frac{r}{4 -r^2}
               \Phi(0)    I_0,
\end{eqnarray}
for $J/\psi$-$\g0$ production. Here $r=4m_c/\sqrt{s}$ and
$C_F = (N_c^2 - 1)/(2 N_c) = 4/3$ is
the color factor. These results are consistent with those in 
Refs.~\cite{brodsky,braaten,chao}.

It is natural to expect the  mixing between the pure states of
$\chi_{c0}$ and $\g0$ if they carry the same quantum numbers
$J^{PC}=0^{++}$ and have small mass difference. The pure
charmonium and glueball states, $\chi_{c0}$ and $\g0$,  
and the mass eigenstates $H_1$, $H_2$ are then linked by
\bea \label{mix} 
\pmatrix{ \chi_{c0}\cr {\cal G}_0\cr} =U \pmatrix{
H_1\cr H_2\cr}, ~~~~~ 
U=\pmatrix{\cos \xi& \sin\xi\cr -\sin \xi&
\cos\xi\cr}, 
\eea 
where $\xi$ is the mixing angle. The mass-squared matrix for $\chi_{c0}$ and 
$\g0$ has the following form and is diagonalized as 
\bea 
U^\dagger \pmatrix{(M_{\chi})^2 & \Delta \cr \Delta& (M_g)^2 \cr} U
= \pmatrix{(M_{1})^2 & 0 \cr 0 & (M_{2})^2 \cr}.
\eea 
We assume that the observed $\chi_{c0}$ state is $H_1$, and hence
$M_1=M(\chi_{c0})_{\rm observed}=3.415~\gev$~\cite{pdg}. Because $H_1$
properties observed in charmonium radiative decays are consistent
with the pure $\chi_{c0}$ assumption, we expect the mixing angle to
be small $\sin^2\xi\leq 0.1$. The mixing angle $\xi$ is real if the 
transition matrix element $\Delta$ is real. If $M_{\chi}\geq M_g$, 
$M_1>M_2$, while if $M_{\chi}\leq M_g$, $M_1<M_2$. The mixing angle 
can be non-negligible ($\sin^2\xi\sim 0.1$) if $|\Delta|\sim |M_{\chi}^2
-M_g^2|$.

The helicity amplitudes for $J/\psi$-$H_1$ and $J/\psi$-$H_2$ productions
keep the forms in Eq. (\ref{hel1}) and the coefficients $a_{\chi,g}$ 
and $b_{\chi,g}$ have to be replaced by $a_{H_1,H_2}$ and $b_{H_1,H_2}$ 
respectively. With Eq. (\ref{mix}), we have 
\bea\label{coe} 
\matrix{
a_{H_1}=a_\chi\cos\xi - a_g\sin\xi,&~~~~~~& 
a_{H_2}=a_\chi\sin\xi + a_g\cos\xi,\cr
b_{H_1}= b_\chi\cos\xi -b_g\sin\xi,&~~~~~~& 
b_{H_2}= b_\chi\sin\xi +b_g\cos\xi. 
}
\eea 
In the $\sin\xi=0$ limit, $H_1$ is a pure $\chi_{c0}$, while $H_2$ is
a pure glueball.

\section{Results}
For $J/\psi$ productions in $e^+ e^-$ annihilation, the mixing effects 
can be measured via the angular distributions of $J/\psi$ and its leptonic 
decays. According to the helicity amplitudes in Eq. (\ref{hel1}),
the triple angular distribution for the process $e^+ e^- \to J/\psi H
\to l^+ l^- H$ ($H$ denotes $H_1$ or $H_2$) is given by
\begin{eqnarray}\label{triple}
\frac{d\sigma}{d\cos\theta d\cos\theta^* d\phi^*} &=&
\frac{ 3 g^4_s e^4 Q^2_c \sqrt{1 - r^2} B}{2048\pi^2s}
\times \Big[ |b_{H}|^2 ( 1 + \cos^2\theta ) (  1 + \cos^2\theta^* )
+ 2 |a_{H}|^2 \sin^2\theta \sin^2\theta^*\nonumber\\
&+&  2 {\re} (a_{H}^* b_{H}) \sin 2\theta
\sin 2\theta^*\cos \phi^*
+ |b_{H}|^2 \sin^2\theta \sin^2\theta^* \cos 2\phi^*
  \Big],
\end{eqnarray}
where $\theta^*$ and $\phi^*$ are the polar and azimuthal angles of
$l^-$ in the $J/\psi$ rest frame. The polar angle $\theta^*$ is measured
from the $J/\psi$ momentum direction in the $e^+e^-$ collision rest frame,
and $\phi^*$ is measured from the scattering plane. In Eq. (\ref{triple}),
$B$ is the branching fraction $B(J/\psi \to l^+ l^-)=5.9\%$ for $l=e,\mu$.

With respect to the mass difference of the mass eigenstates $H_1$ and $H_2$,
we meet with two different cases:
(1) If the mass difference is large enough, only $H_1$ 
has been identified as the $\chi_{c0}$ resonance. 
(2) It is also possible that the mass difference is quite small and both 
$H_1$ and $H_2$ contribute to the observed resonance peak.
Therefore we will consider these two cases separately.

In the first case, we rename $H_1$ as $H$. By integrating over $\theta^*$ 
and $\phi^*$ and comparing with Eq. (\ref{angular}), we obtain
\bea \label{alpha}
\alpha=\frac{b_H^2-a_H^2}{b_H^2+a_H^2}
=\frac{\sigma_T-2\sigma_L}{\sigma_T+2\sigma_L},
\eea
where $\sigma_T$ and $\sigma_L$ are the cross sections for the transversely 
and longitudinally polarized $J/\psi$ productions, and 
$\sigma=\sigma_T+\sigma_L$ is the total cross section. One can obtain 
the $\cos\theta^*$ distribution by integrating over $\theta$ and $\phi^*$
in Eq. (\ref{triple}), and obtain $d\sigma/d\cos\theta^*\sim 1+
\alpha^*\cos^2\theta^*$.  It is easy to find $\alpha^*=\alpha$ according 
to Eq. (\ref{triple}). The interference term $2{\re}(a_H^* b_H)$ can be 
measured through the combination of the $\theta$, $\theta^*$ and $\phi^*$ 
distributions. We introduce a new quantity $\alpha_{\rm off}$ as the 
normalized interference term
\bea \label{aoff}
\alpha_{\rm off}=\frac{2{\re}(a_H^* b_H)}{b_H^2+a_H^2}.
\eea
Both $\alpha$ and $\aoff$ are bounded in the interval $[-1,1]$.

Now we present our numerical results based on the input parameter values
$e^2/4\pi=1/129.6$, $g_s^2/4\pi=0.26$, $m_c=1.5\gev$, and 
$\sqrt{s}=10.58\gev$.
The value of the $J/\psi$ wave function at the origin can be obtained 
from $\Gamma(J/\psi\to e^+ e^-)$, $|\Phi(0)|^2=0.0336\gev^{3}$
in the leading order of NRQCD~\cite{braaten, hagiwara, hagiwara2}. 

The $\chi_{c0}$ wave function at the origin $\Phi'(0)$ is estimated 
from the observed width $\Gamma(\chi_{c0}\to\gamma\gamma)$~\cite{braaten}. 
Since we assume that the observed state is a mixture $H$, the partial 
width is actually proportional to $|\cos\xi \Phi'(0)|^2$. Here we assume 
that only the charmonium content of $H$ contributes to the decay 
$\chi_{c0}\to \gamma\gamma$ at the leading order in $\alpha_s$. Therefore 
we obtain $|\cos\xi \Phi'(0)|^2=0.0117\gev^{5}$. It should be noted that 
the partial widths and the observed radiative transition rates can be
affected at several percent level for the mixing angle of the order
of $\sin^2\xi\sim 0.1$.

The upper bound of the glueball wave function $|I_0|$ is obtained
from the radiative $\Upsilon$ decay. The CUSB Collaboration reported 
a 90\%-C.L upper limit for the branching ratio $Br(\Upsilon \to \gamma H)$ 
which is about 0.01\%-0.15\% for the scalar boson ($H$) mass between 
2 and 8.5 GeV~\cite{cusb}. The upper bound is obtained in terms of 
the CUSB mass resolution, 20 MeV. As discussed in Ref.~\cite{brodsky},
if the $H$ decay width $\Gamma_H$ is larger than the resolution, 
the upper bound should be loosen by a factor of $\Gamma_H/(20\rm MeV)$.
In Ref.~\cite{brodsky}, the authors considered the pure glueball 
production process, $e^+e^- \to J/\psi \g0$, and they assumed the glueball 
decay width should be less than about 100 MeV, which is the full width 
at half maximum of the `$\chi_{c0}$' peak in the Belle fit to the $J/\psi$
recoil mass distribution. They found the upper bound
for the glueball wave function $|I_0|^2<5.8\times 10^{-3}~{\rm GeV}^2$.
Accordingly, the rate of the cross sections of the pure glueball and
charmonium productions $\sigma_{J/\psi \g0}/\sigma_{J/\psi \chi_{c0}}$
which is proportional to $|I_0|^2/|\Phi'(0)|^2$, is less than 0.72.
Similar upper bound applies in our case, except that
the radiative $\Upsilon$ decay rate constrains 
$|\sin\xi I_0|^2$ instead of $|I_0|^2$, since $\Upsilon$ can decay 
to the glueball content of $H$ only, at the leading order in $\alpha_s$.
In the $e^+e^-\to J/\psi H$ process, because the contributions from the 
charmonium and glueball contents are proportional to $\cos^2\xi$ and 
$\sin^2\xi$, respectively, the upper bound of Ref.~\cite{brodsky} leads 
to the bound $(\sin^2\xi\sigma_{J/\psi \g0})/(\cos^2\xi\sigma_{J/\psi 
\chi_{c0}})<0.72$. We therefore define the following quantity 
\bea\label{r}
{\cal R}={\rm Sign}\left[\sin\xi \frac{I_0}{\Phi'(0)}\right] \times
\frac{\sin^2\xi~\sigma_{J/\psi \g0}}{\cos^2\xi~\sigma_{J/\psi \chi_{c0}}}
\eea
to present our numerical results. Here ${\rm Sign}\left[\sin\xi \frac{I_0}
{\Phi'(0)}\right]$ determines the sign of the interference terms. The 
$\xi$-dependence of our results enters only through the rate $\r$. For 
simplicity, we assume that the wave functions $\Phi'(0)$ and $I_0$ are real. 

The magnitude of the parameter $\r$ can be as large as unity if $H_1$ is 
the only gluon-rich state which contribute to the radiative $\Upsilon$ 
decay. It is, however, unlikely that the $H_2$ mass lies outside of the
reach of the CUSB search \cite{cusb}. If the $H_2$ mass is in the region 
2-8.5 GeV, the radiative $\Upsilon$ decay rate should constrain 
$|\cos\xi I_0|^2$, and therefore gives an upper bound on $|\r|$ in Eq. 
(\ref{r}) of the order of $\tan^2\xi(\Gamma_{H_2}/100\mev)\times 0.72$. 
If $\sin^2\xi<0.1$ and if the $H_2$ width $\Gamma_{H_2}$ is below 500 
MeV, the allowed range of $\r$ is reduced to $|\r|\leq 0.4$.

\begin{figure}
\centerline{\epsfysize 2.0 truein \epsfbox{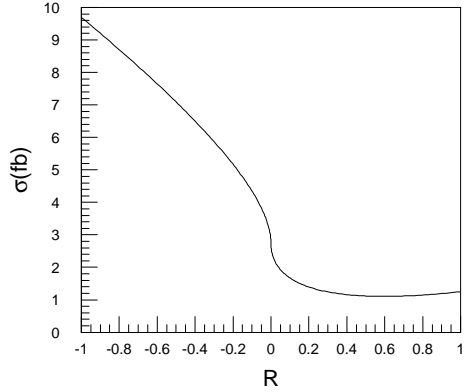}}
\caption{
The cross section for $e^+ e^- \to J/\psi H$ in the interval
$-1<{\cal R}<1$ for the case that only one of the mass eigenstates
lies in the `$\chi_{c0}$' resonance.
}
\end{figure}
In Fig. 2, we show the cross section for $e^+ e^- \to J/\psi H$ in 
the interval $-1<{\cal R}<1$. We find the cross section from the 
contribution of the $\chi_{c0}$ content, $\cos^2\xi~\sigma_{J/\psi \chi_{c0}}$, 
is 2.7 fb, much smaller than $12\pm 4$ fb which we estimate
from the preliminary Belle data~\cite{belle3}.
We note that the contribution from the interference term can significantly 
increase the cross section in the region $-1<{\cal R}<0$, and decrease 
the cross section for $0<{\cal R}<1$. For instance, at ${\cal R}=-0.1$, 
the total cross section is about 4.3 fb, while $\sin^2\xi~\sigma_{J/\psi \g0}=
|{\cal R}|\cos^2\xi~\sigma_{J/\psi \chi_{c0}}=0.27$ fb. The interference term 
makes up the difference $4.3-2.7-0.27=1.3$ fb of the total cross section, 
which is about five times greater than the direct contribution from 
the glueball content. At ${\cal R}=-0.4$, the total cross section is about
6.5 fb among which 2.7 fb comes from the interference contribution.
While at ${\cal R}=-1$, the total cross section can be as large as 9.6 fb
(4.2 fb from the interference contribution), which
is comparable with the central value of the experimental data.

In Fig. 3, we plot the angular asymmetry parameters $\alpha$ (solid line) 
and $\aoff$ (dashed line) which are defined in Eqs. (\ref{alpha}) and 
(\ref{aoff}) as functions of ${\cal R}$. $\alpha$ for the charmonium 
content is 0.25 at ${\cal R}=0$, while for the glueball content is $-$0.85. 
The Belle fit finds $\alpha=-1.1^{+0.8}_{-0.6}$, consistent with $-$1
\cite{belle2}. Although the error is still quite large, we may state that 
$\alpha <0$ is favored by the
Belle data~\cite{belle2}. In Fig. 3, $\alpha <0$ corresponds to the region 
$\r<-0.045$ and $\r>0.85$. At $\r=-0.4$ and $\r=-1$, $\alpha$ can be as 
small as $-0.30$ and $-0.43$, respectively. 
The region $\r>0.85$ is not favored because of the suppressed cross section
in Fig. 2. For the case of the asymmetry parameter $\aoff$, $\aoff=-0.97$ 
for the charmonium content ($\r=0$), and $\aoff=-0.52$
for the glueball content (${\cal R}\to \pm \infty$). According to
Eqs. (\ref{alpha}) and (\ref{aoff}), $\aoff=\pm 1$ should correspond to
$\alpha=0$ which reflects the produced $J/\psi$ mesons 
are unpolarized ($\sigma_T=2\sigma_L$). $\aoff=0$ corresponds to $\alpha=1$ 
or $-1$, that is, all $J/\psi$ are transversely or longitudinally polarized.
These can be observed in Fig. 3 at $\r=-0.045$ ($\alpha =0$ while $\aoff=-1$),
$\r=0.27$ ($\alpha =1$ while $\aoff=0$), and $\r=0.85$ 
($\alpha =0$ while $\aoff=1$). In the region $-1<\r<-0.045$ favored by
the Belle data, $\aoff$ varies slightly from $-1$ to $-0.90$.
\begin{figure}
\centerline{\epsfysize 2.0 truein \epsfbox{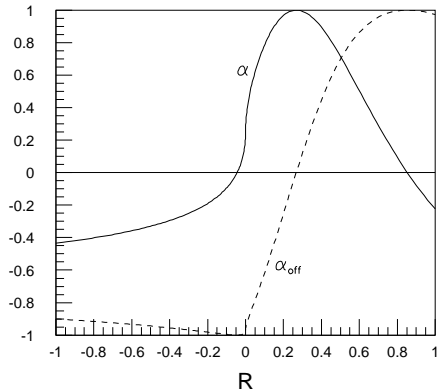}}
\caption{
The asymmetries $\alpha$ and $\alpha_{\rm off}$ in the interval
$-1<{\cal R}<1$ for the case that only one of the mass eigenstates
lies in the `$\chi_{c0}$' resonance.
}
\end{figure}

Now we turn to the second case in which both of the two mass eigenstates
$H_1$ and $H_2$ are hidden in the `$\chi_{c0}$' resonance peak of the
$J/\psi$ recoil mass distribution in the Belle data~\cite{belle2}. In 
this case, one can obtain the total cross section and asymmetries by 
summing over the triple angular distributions in Eq.~(\ref{triple}) for 
$H_1$ and $H_2$. By using the relation between the coefficients 
$a_{H_1,H_2},b_{H_1,H_2}$ and $a_{\chi,g},b_{\chi,g}$ in Eq. (\ref{coe}), 
we find that 
\bea
a_{H_1}^2+ a_{H_2}^2&=&a_{\chi}^2+a_g^2, \nonumber \\
b_{H_1}^2+ b_{H_2}^2&=&b_{\chi}^2+b_g^2,  \nonumber \\
a_{H_1}^*b_{H_1}+ a_{H_2}^*b_{H_2}&=&a_{\chi}^*b_{\chi}+a_g^* b_g.
\eea
From the above relations, we can find that the sum of the contributions from 
the two mixed states $H_1$ and $H_2$ is essentially identical to the sum of 
the contributions from the pure $\chi_{c0}$ and $\g0$ states.

The sum of the cross sections for $J/\psi$-$\chi_{c0}$ and $J/\psi$-$\g0$ 
productions is simply to be $\sigma_{J/\psi\chi_{c0}}+\sigma_{J/\psi\g0}
=(1+{\bar\r})\sigma_{J/\psi\chi_{c0}}$, where
\bea
{\bar\r}=\frac{\sigma_{J/\psi \g0}}{\sigma_{J/\psi \chi_{c0}}}.
\eea
For the leading order estimate of $\sigma_{J/\psi\chi_{c0}}=2.7$ fb, 
the sum of the cross sections can be as large as 5.4 fb for ${\bar\r}=1$, 
which is to be compared with our estimate of $12\pm 4$ fb~\cite{belle3}.

\begin{figure}
\centerline{\epsfysize 2.0 truein \epsfbox{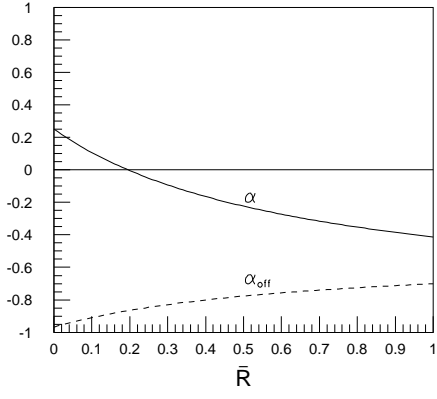}}
\caption{
The asymmetries $\alpha$ and $\alpha_{\rm off}$ in the interval
${\bar {\cal R}}=\sigma_{J/\psi \g0}/\sigma_{J/\psi \chi_{c0}}<1$ 
for the case that both of the mass eigenstates
$H_1$ and $H_2$ lie in the `$\chi_{c0}$' resonance.
}
\end{figure}
The asymmetry $\alpha$ for this case is given by
\bea \label{alphanon}
\alpha=\frac{b_\chi^2-a_\chi^2+b_g^2-a_g^2}{b_\chi^2+a_\chi^2+b_g^2+a_g^2},
\eea
and $\aoff$ is expressed as
\bea \label{anonoff}
\alpha_{\rm off}=\frac{2{\re}(a_\chi^* b_\chi+a_g^* b_g)}
{b_\chi^2+a_\chi^2+b_g^2+a_g^2},
\eea
where $a_{\chi,g}$ and $b_{\chi,g}$ follow from Eqs. (\ref{abchi}) and 
(\ref{abg}). They are plotted against ${\bar \r}$ in Fig. 4. One can find 
$\alpha=0.25$ and $\aoff=-0.97$ by setting ${\bar \r}=0$ for the $J/\psi$-
$\chi_{c0}$ production, and $\alpha=-0.85$ and $\aoff=-0.52$ for the 
$J/\psi$-$\g0$ production in the ${\bar \r}\to \pm \infty$ limit, which are 
the same as those for the previous case. $\alpha$ falls off and $\aoff$ 
goes up with increasing ${\bar\r}$. ${\bar\r}>0.20$ gives $\alpha<0$ and 
$\aoff>-0.87$. At ${\bar\r}=1$, $\alpha=-0.41$ and $\aoff=-0.70$.

In this paper, we studied the disagreement between the experiment and 
the NRQCD prediction for the $J/\psi$-$\chi_{c0}$ associated production,
in the production cross section and the $J/\psi$ angular distribution.
We introduced a charmonium-glueball mixing mechanism to explain this 
discrepancy and propose to study various angular distributions to
explore the possible mixing effects. Our results may facilitate the 
present and future experimental measurements to resolve possible glueball 
contents in the charmonium resonances.


\section*{Acknowledgments}
The work of SD is supported in part by National Natural Science Foundation 
of China under Grant No. 90103012 and by the Core University Exchange Program
between Chinese Academy of Sciences and the Japan Society for the Promotion 
of Science (JSPS). The works of KH and ZHL are supported in part by 
Grant-in-Aid Scientific Research from MEXT, Ministry of Education, 
Culture, Science and Technology of Japan. The work of ZHL is supported 
in part by JSPS.


\begin{thebibliography}{9}
\bibitem{belle3}
Ref.~\cite{belle2} presents the accumulated event numbers 
for the $J/\psi$-$\eta_c$ and $J/\psi$-$\chi_{c0}$ productions are 
$175\pm 23$ and $61\pm 21$ respectively. With the published result for 
the $J/\psi$-$\eta_c$ cross section $\sigma(e^+e^-\to J/\psi \eta_c)
\geq 33\pm 11$ fb~\cite{belle, belle2}, we estimate the $J/\psi$-$\chi_{c0}$ 
cross section to be about $12\pm 4$ fb, provided that the detection 
efficiencies do not differ much between the two channels.
\end{thebibliography}
\end{document}